\documentclass[twocolumn,amsmath,amssymb,prb,superscriptaddress,showpacs]{revtex4}
\usepackage{graphicx}
\begin{document}
\preprint{Preprint Number}
\title{Neutron scattering experiments and simulations near
the magnetic percolation threshold of $\rm{Fe}_x\rm{Zn}_{1-x}\rm{F}_2$}
\author{G.~\'Alvarez}
\affiliation{Departamento de F\'{\i}sica Te\'orica II (M\'etodos Matem\'aticos
  de la F\'{\i}sica), Facultad de Ciencias F\'{\i}sicas, 
  Universidad Complutense, 28040 Madrid, Spain.}
\author{N.~Aso}
\affiliation{Department of Physics and Earth Sciences, Faculty of Science, University of
the Ryukyus,
Senbaru 1, Nishihara, Okinawa 903-0213, Japan.}
\author{D.~P.~Belanger}
\author{A.~M.~Durand}
\affiliation{Department of Physics,University of California,
Santa Cruz, CA 95064, USA.}
\author{V.~Mart\'{\i}n-Mayor}
\affiliation{Departamento de F\'\i{}sica Te\'orica I, Universidad
  Complutense, 28040 Madrid, Spain.}
\affiliation{Instituto de Biocomputaci\'on y F\'{\i}sica de Sistemas
  Complejos (BIFI), Spain.}
\author{K.~Motoya}
\affiliation{Department of Physics, Faculty of Science and Technology,Tokyo University of Science, Noda, Chiba 278-8510, Japan.}
\author{Y. Muro}
\affiliation{Department of Physics, Faculty of Science and Technology,Tokyo University of Science, Noda, Chiba 278-8510, Japan.}
\affiliation{Liberal Arts and Sciences, Faculty of Engineering, Toyama Prefectural
University, Kurokawa 5180, Imizu, Toyama 939-0398, Japan.}
\date{\today}
\begin{abstract}
The low temperature excitations in the anisotropic antiferromagnetic
$\rm Fe_{1-x}Zn_xF_2$ for $x=0.25$ and $0.31$, at and just above
the magnetic percolation threshold concentration $x_{\mathrm p}=0.25$, were measured using
inelastic neutron scattering.  The excitations were
simulated for $x=0.31$ using a localized, classical excitation model, which accounts
well for the energies and relative intensities of the excitations
observed in the scattering experiments. 
\end{abstract}
\maketitle
\section{INTRODUCTION}
In two and three dimensions, spin wave excitations
are well studied in pure isotropic and anisotropic
insulating antiferromagnets \cite{HU70,NLD69,BJR78,ih78,ws66}.  Magnetic
excitations are significantly modified by magnetic dilution
introduced by site substitution of the magnetic ions with
diamagnetic ones.  For isotropic systems near the magnetic percolation
threshold concentration, $x_{\mathrm p}$, both well-resolved,
local spin excitations as well
as crossover from spin wave excitations to fracton excitations
on a fractal-like lattice \cite{IIYHS98,ITFN98,INKA09,INA11, IFNTI94,nyo94}, have been
characterized near $x_{\mathrm p}$ in two and three dimensions.
Local spin excitations have been observed in dilute two-dimensional
anisotropic systems near $x_{\mathrm p}$ \cite{io92}.
In three dimensions, magnetic excitations have been studied for
the magnetically dilute anisotropic systems
$\rm Mn_{0.5}Zn_{0.5}F_2$ \cite{UB87} and
$\rm Fe_xZn_{1-x}F_2$ \cite{PA94,SA02,RABNYF07}.

The parent compounds $\rm MnF_2$
and $\rm FeF_2$ exhibit comparable exchange energies
and corresponding spin wave dispersions, but the
$\rm FeF_2$ system has an order of magnitude
larger anisotropy and a correspondingly larger
spin wave gap.  The excitations in the $\rm Mn_{0.5}Zn_{0.5}F_2$
system were interpreted in terms of spin wave to fracton
crossover as the scattering wavevector $q$ increases.
The behavior of the $\rm Fe_xZn_{1-x}F_2$, for $x \ge 0.31$,
has been interpreted \cite{PA94} as showing both spin wave
and local spin excitations for small $q$ and local spin excitations
for large $q$.  In this study, we examine magnetic excitations
with high resolution neutron scattering experiments and computer
simulations in $\rm Fe_xZn_{1-x}F_2$ as $x$ approaches $x_{\mathrm p}=0.25$.

The well-characterized random-exchange 
antiferromagnet $\rm Fe_xZn_{1-x}F_2$,
with its simple structure and interactions, is
an ideal anisotropic three-dimensional ($d=3$) system in which to study magnetic
excitations through inelastic neutron scattering measurements
and theoretical modeling and simulation.
Magnetic excitations in the $d=3$ anisotropic
antiferromagnet $\rm FeF_2$ have been very well characterized.\cite{HU68,HU70}
The structure of $\rm FeF_2$ and diamagnetic $\rm ZnF_2$
are similar.\cite{SR54}  The antiferromagnetic spins in $\rm FeF_2$ form
a tetragonal lattice with two interpenetrating
sublattices. The dominant antiferromagnetic inter-sublattice
$J_2$ exchange interaction is
between the body-center and body-corner magnetic ions.
The intra-sublattice ferromagnetic $J_1$ and
frustrating antiferromagnetic $J_3$
exchange interactions are much smaller  
(cf.~Sec.~III for further details).
Best fit values from inelastic neutron scattering
measurements are shown in Table I.  $\rm FeF_2$ and
diamagnetic $\rm ZnF_2$ mix well during crystal growth
to form $\rm Fe_xZn_{1-x}F_2$, which is a dilute,
anisotropic, three-dimensional antiferromagnet.  The
occupation of sites by $\rm Fe^{2+}$ ions with
$S=2$ or diamagnetic $\rm Zn^{2+}$ ions appears close
to random, though slight clustering cannot be ruled out.
It appears that $J_2$ does not vary significantly
with dilution.\cite{KJSMD81,A80}  There is limited information
about the effect of dilution on the anisotropy, but it
also does not appear to vary by a large amount.\cite{A80}

Magnetic ordering in $\rm Fe_xZn_{1-x}F_2$
has been experimentally studied previously
\cite{BMMKJE91,ACVAMMRC91,PA94,SI97,JDNB97,BB00,SA02,BRC03,BRC05,RABNYF07,BY07,LBBRC12} at magnetic
concentrations $x$ equal to or near
$x_{\mathrm p} = 0.246$, the magnetic percolation
threshold for the body-centered tetragonal magnetic
structure with an interaction between the body-centered
and corner ions ($J_2$ in $\rm Fe_xZn_{1-x}F_2$).
The $H=0$ random-exchange transition should be expected for
$x>x_{\mathrm p}$ if there is only the dominant exchange interaction $J_2$.
However, the small $J_1$ and $J_3$ interactions in
$\rm Fe_xZn_{1-x}F_2$ could become influential near $x_{\mathrm p}$.
The prior experiments in zero-field have demonstrated
that for concentrations $x \le 0.31$ \cite{BMMKJE91} 
there is, at best, very weak long-range antiferromagnetic order at
low temperatures.  The system exhibits spin-glass-like behavior,
dominated by slow dynamics near the percolation threshold, possibly a
result of the frustrating $J_3$ interaction.

Early inelastic neutron scattering measurements in $\rm Fe_xZn_{1-x}F_2$
were compared to a simple treatment with the excitation energies
assigned \cite{PA94} as $(z/n)E(q)$, where $n=8$ represents the number of
neighbors in $\rm FeF_2$, $z \le n$ is the possible number of neighbors
of a given spin in the magnetically dilute system, and $E(q)$ is the spin wave
energy as a function of the scattering wave vector $q$ in $\rm Fe_xZn_{1-x}F_2$.
The intensities are assigned by
the combinatorial probabilities of finding $z$ neighbors of a given spin.  While giving
a fairly accurate description of the overall spread in energy, this
description fails in the detailed structure of the excitation spectrum
when higher energy resolution measurements resolve individual peaks.
Similar results were found in far-infrared absorption experiments,
high magnetic field pulsed laser absorption, and inelastic
neutron scattering experiments for $x \ge0.4$.\cite{PA94,SA02,RABNYF07}  It was observed in the
pulsed laser absorption measurements that the peaks become more easily
resolvable for $x=0.4$ and that for this case the simplistic modeling
described above proves wholly
inadequate;\cite{RABNYF07} the spacings of the resolved peaks do not
correspond to the simple model.  The excitations were found to
be largely localized, having little dispersion.

Here, we present a high resolution neutron-scattering study of $\rm
Fe_xZn_{1-x}F_2$ close to its percolation threshold. In agreement with the
aforementioned work, the excitations show little or no dispersion. We show
that a model of localized excitations accounts for our spectra even
quantitatively.

The layout of the remaining part of this paper is as follows. In
Sec.~\ref{sect:experiments}, we report our new experimental results.  The
obtained spectra are rationalized through a simple model in
Sec.~\ref{sect:model}. Finally, we present our conclusions in
Sec.~\ref{sect:discussion}.

\section{Experimental results}\label{sect:experiments}

\begin{figure}
 \includegraphics[width=9cm]{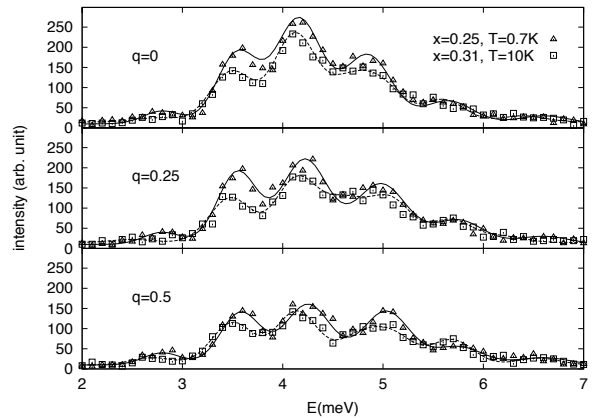}
 \caption{\label{excite} Experimental intensities vs energy
for $x=0.25$ and $0.31$
for the zone center, $q=0$, for $q=0.25$, and for the zone boundary, $q=0.5$.  
The data for $x=0.25$ were taken at $T=0.7$K and those at $x=0.31$ were
taken at $T=10$K.  The curves are guides to the eye constructed from
Gaussian peaks.  The Gaussian peak locations are the same for each value of $q$.
The ratio of intensities for $x=0.25$ and $x=0.31$ is arbitrary.} 
\end{figure}

\begin{figure}
 \includegraphics[width=9cm]{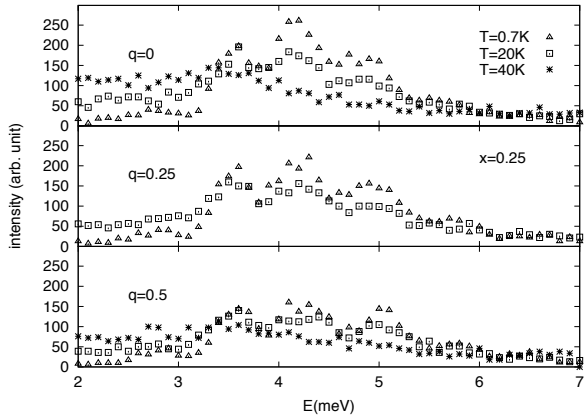}
 \caption{\label{excite_T} Experimental intensities vs energy 
for $x=0.25$ at $T=0.7$, $20$, and $40$K to show the
temperature dependence of the neutron scattering spectra.} 
\end{figure}

The results at $x=0.4$ motivated experiments closer to the percolation
threshold and we have conducted inelastic neutron scattering studies for
$x=0.25$ and $x=0.31$.  Neutron scattering measurements were carried out using
the high energy-resolution triple-axis spectrometer C1-1 installed at the
JRR-3M reactor of JAEA in Tokai operating with a horizontally focusing
analyzer with a final neutron energy of $E_f=3.1$ meV.  The energy resolution
at the elastic position is 0.09 meV (full width at half maximum) but it
increases to 0.46 meV as the energy transfer increases to 8 meV.
Single-crystal samples were mounted in a closed-cycle refrigerator with the
c-axis perpendicular to the scattering plane.  We examined a $\rm
Fe_{0.25}Zn_{0.75}F_2$ single crystal with a mass of $2.24$g and a $\rm
Fe_{0.31}Zn_{0.69}F_2$ single crystal with a mass of $1.72$g.  The magnetic
concentrations of the optical-quality crystals were determined using density
measurements.  The resulting scattering spectra of the experiments, shown in
Figs.~\ref{excite} and~\ref{excite_T}, indicate localized excitations since
the excitation energies are largely independent of the scattering wave vector
$q$. The simple approximation described above yields peaks similar to the
resolved experimental peaks in the spectra, but the energies are not well
predicted, as discussed below.  It was clear that a more realistic calculation
was needed to describe the peaks and to elucidate what governs the details of
the $E$ vs $q$ spectra.  We discuss simulations below that capture essential
characteristics of the experimental results.

\section{Modeling the spectra}\label{sect:model}

\begin{figure}
\includegraphics[width=9cm]{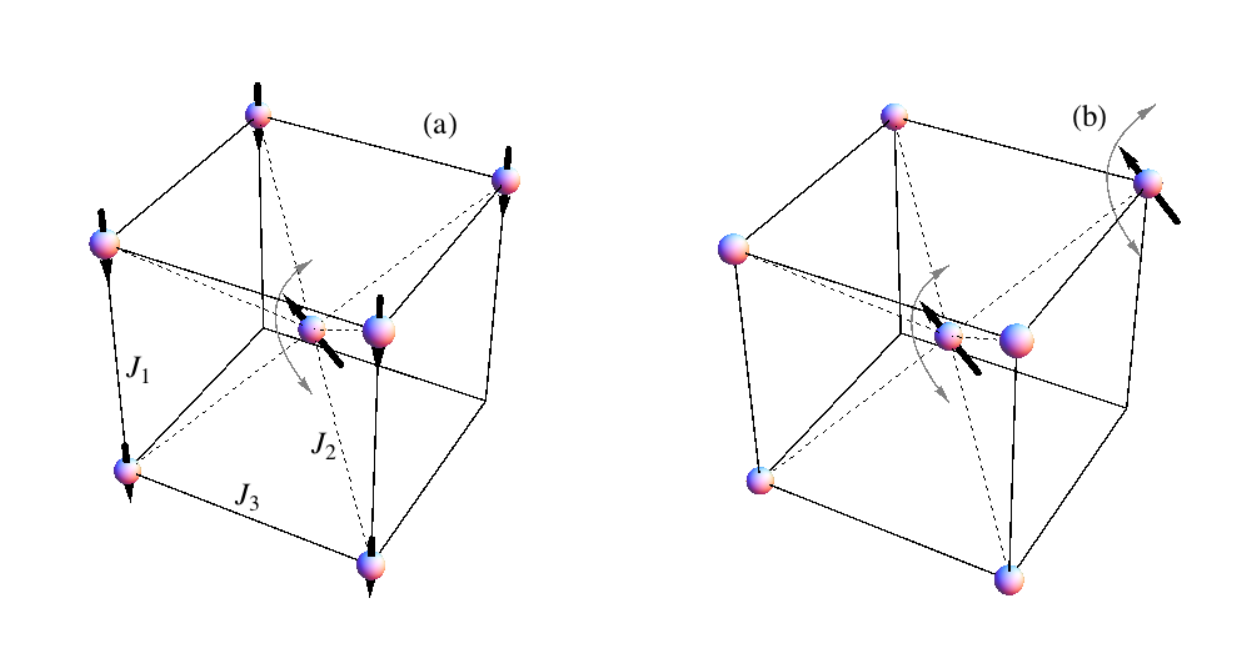}
 \caption{\label{unitcells} (a) Unit cell with seven occupied sites of which the central site is active and the
 six dimmed sites are held frozen in an antiferromagnetic configuration. (b) The same unit cell with all
 the occupied sites being active (for clarity, only two of them are marked with flipping arrows).
 The corresponding frozen environment is the next shell of neighbors
 (not shown in the figure).}
\end{figure}

We have remarked in the previous section that, to a large extent, the scattering spectra
are independent of the scattering wave vector, which suggests that the underlying
excitations are spatially localized and therefore can be described by a local model.
Local spin Hamiltonians for pure $\rm{Fe}\rm{F}_2$ have been extensively
discussed,\cite{HU68,HU70,HU72} and include both an on-site interaction characterized
by an anisotropy parameter $D$, and Heisenberg exchange interactions characterized by three coupling
constants $J_1$, $J_2$ and $J_3$. The strongest exchange is the antiferromagnetic
$J_2$ that couples nearest neighbors  belonging to different sublattices of the
body-centered tetragonal magnetic lattice of $\rm{Fe}$ atoms,
while $J_1$ and $J_3$ couple atoms belonging to
the same sublattice, i.e., bonds parallel to the edges of the unit cell (Fig.~\ref{unitcells}).

Following the same pattern, we model the disordered sample with the following Hamiltonian:
\begin{eqnarray}
  H & = & - D \sum_{i} \epsilon_i \big[S_i^{(z)}\big]^2
               + J_2 \sum_{\langle i,j\rangle} \epsilon_i \epsilon_j {\boldsymbol  S}_i\cdot{\boldsymbol S}_j
               \nonumber\\  
      &  & {} + J_1\sum_{\langle\langle i,j\rangle\rangle} 
                          \epsilon_i \epsilon_j {\boldsymbol S}_i\cdot{\boldsymbol S}_j
                 +J_3\sum_{\langle\langle\langle i,j\rangle\rangle\rangle}\epsilon_i \epsilon_j
                          {\boldsymbol S}_i\cdot{\boldsymbol S}_j,
   \label{eq:hamil}
\end{eqnarray}
where the ${\boldsymbol S}_i$ are $S=2$ spin operators that represent the
$\rm{Fe}$ atoms, and where the substitutional disorder is represented by statistically
independent random variables $\epsilon_i$ which take the value one with probability
$x$ (the fraction of $\rm{Fe}$ atoms) and the value zero with probability $1-x$
(the fraction of nonmagnetic $\rm{Zn}$ atoms symbolized as empty sites in Fig.~\ref{unitcells}).

\begin{table}
  \caption{Best fit parameters, taken from Ref.~\onlinecite{HU70}, for the
    spin Hamiltonian of pure $\rm{Fe}\rm{F}_2$.\label{tab:param}}
  \begin{ruledtabular}
  \begin{tabular}{cccc}
  $D (\rm{cm}^{-1})$ & $J_1 (\rm{cm}^{-1})$ & $J_2 (\rm{cm}^{-1})$ & $J_3 (\rm{cm}^{-1})$ \\
  \hline
  $6.46 (+0.29,-0.10)$ & $-0.048\pm 0.060$ & $3.64\pm0.10$ & $0.194\pm0.060$\\
  \end{tabular}
  \end{ruledtabular}
\end{table}

\begin{figure}
 \includegraphics[width=9cm]{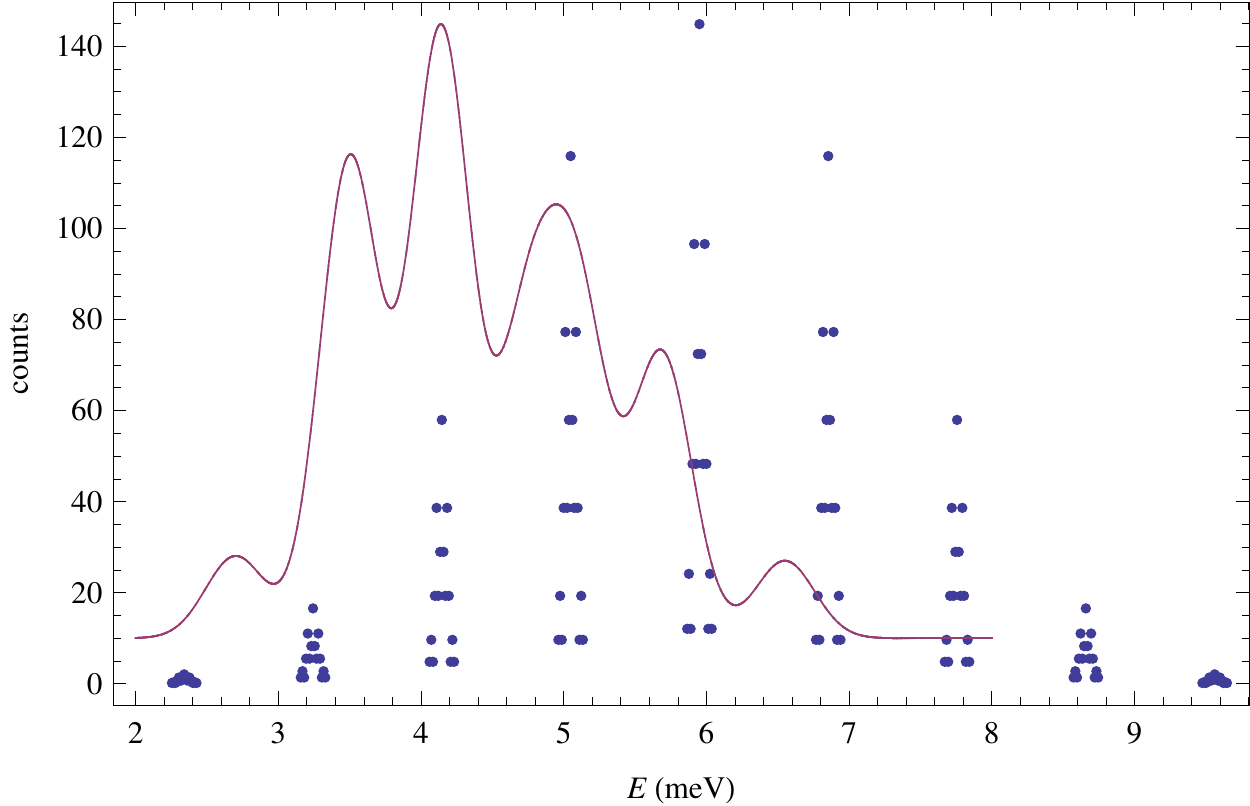}
 \caption{\label{formula} Semiclassical approximation to the spectrum provided
 by~Eq.~\eqref{eq:formula} averaged over the respective number of neighboring sites
 with the pure sample parameters.~\cite{HU70}
The solid line is a fit to the experimental data of Fig.~\ref{excite} corresponding
to $x=0.31$ and $q=0.5$, and suggests that the pure sample parameters need to
be modified to describe the highly diluted sample.}
\end{figure}

Table~\ref{tab:param} shows the parameters obtained by Hutchings~\emph{et al.}~\cite{HU70}
by fitting inelastic neutron scattering data of $\rm{Fe}\rm{F}_2$ to the spin Hamiltonian, and
it is often assumed that the same values can be used for the analysis of the dilute
antiferromagnet~\cite{PA94,SA02} (we will elaborate on this point later).
The fact that  $|J_1|, J_3 \ll J_2 < D$ allows us to get an
estimate of the spectrum at low temperatures by ignoring in the Hamiltonian~\eqref{eq:hamil}
the terms proportional to $J_1$ and $J_3$ and making the approximation
${\boldsymbol S}_i\cdot{\boldsymbol S}_j\approx S_i^{(z)}S_j^{(z)}$
in the terms proportional to $J_2$. For the sake of definiteness let us consider a site $i$ on
the A sublattice, with $n_2$ $\rm{Fe}$ neighbors (the probability distribution function for
$n_2=0,1,\ldots,8$ is binomial). Since at very low temperatures the magnetic state of the sample is
essentially the N\'eel state ($S_i^{(z)}=+2$ if the site $i$ belongs to the A sublattice and
$S_i^{(z)}=-2$ if $i$ is in the B sublattice), the local magnetic field felt by the spin $i$ due to 
the $n_2$ surrounding atoms is $-2n_2J_2$. Hence, the contribution of spin $i$ to the energy is
$E_i=-D[S_i^{(z)}\big]^2-2n_2J_2S_i^{(z)}$.  An incoming neutron typically
causes a spin flip $S_i^{(z)}=2\rightarrow S_i^{(z)}=1$; the energy of
such a transition is
\begin{equation}
	\label{eq:e}
	\Delta E \approx 3 D + 2 n_2 J_2,
\end{equation}
and therefore to a first approximation the spectrum consists of nine evenly spaced
zero-width peaks with a binomial distribution of intensities. The anisotropy parameter $D$
determines the average position of the peaks, while the spacing between peaks is proportional
to $J_2$. Actually, a more accurate description can be obtained by averaging the immediate
generalization of Eq.~\eqref{eq:e}
\begin{equation}
	\label{eq:formula}
	\Delta E \approx 3 D - 2 n_1 J_1 + 2 n_2 J_2 - 2 n_3 J_3,
\end{equation}
over the respective number of neighboring sites $0\le n_1\le 2$, $0\le n_2\le 8$, $0\le n_3\le 4$.
In Fig.~\ref{formula} we show the result of this calculation with the parameters of the pure
sample~\cite{HU70} and (zero-width) intensities proportional to the products of the respective combinatorial
weights. This figure shows how the coupling constants $J_1$ and $J_3$ contribute to the effective
spread of the peaks. In fact, comparison with the fit to the experimental data suggests
that all the pure sample parameters need to be modified to describe the highly diluted sample and,
in particular, that the anisotropy parameter $D$ is too large. As we will see
in the forthcoming discussion, the parameters that provide the best fit to the experimental
data depend on the approximation scheme used to study the model~Eq.~\eqref{eq:hamil}.
However, the semiclassical picture given by Eq.~\eqref{eq:formula}
and illustrated in Fig.~\ref{formula} remains qualitatively correct in the full simulation.

Our approach to simulate the experimental spectra is a two-step procedure.
In the first step we use the full Hamiltonian~\eqref{eq:hamil} but maintain the
approximation 
${\boldsymbol S}_i\cdot{\boldsymbol S}_j\approx S_i^{(z)}S_j^{(z)}$
for the exchange terms, so that we can generate typical local environments
by a classical Monte Carlo simulation. More concretely, we generate
equilibrium spin configurations at $T=10~\rm{K}$ using a heat bath combined
with a cluster method in lattices with $2\times 32^3$ sites with an $\rm{Fe}$
density $x=0.31$. After we equilibrate ten such lattices (samples) we pick at
random on each sample 1000 nonempty sites. We call this site, together with
its $n$ nearest neighbors, as illustrated in Fig.~\ref{unitcells}(b), a dynamic shell,
and the next shell of atoms [not shown in Fig.~\ref{unitcells}(b)], the local environment.
Figure~\ref{unitcells}(a) illustrates a simpler version of this idea, in which there
is only one spin in the dynamic shell (the central spin marked by a flipping arrow)
and there are six frozen atoms in an antiferromagnetic state that constitute
the local environment.

In the second step of our procedure we use the full Hamiltonian~\eqref{eq:hamil}
with full quantum spin operators for each atom on the dynamic shell, while the
nondynamic spins that constitute the local environment are kept fixed
and act in effect as boundary conditions for the dynamic shell.
Since the third component of the total spin for each dynamic shell
\begin{equation}
  S^{(z)} = \sum_{i\ \text{dynamic}} S_i^{(z)}
\end{equation}  
commutes with the Hamiltonian, we find the ground state (G.S.) within the subspace
corresponding to the N\'eel state $S^{(z)}=S_\text{N\'eel}$, which amounts
to diagonalizing a square matrix with up to $3000$ states, and the excited
states by diagonalizing the Hamiltonian restricted to the subspaces
$S^{(z)}=S_\text{N\'eel}\pm 1$ allowed by the selection rules.
The transition energies are simply the energy differences
\begin{equation}
  \Delta E^\pm_k = E_k^{S_\text{N\'eel}\pm 1}
                              -
                              E_\text{G.S.}^{S_\text{N\'eel}}.
  \label{eq:trans}
\end{equation}

The exact calculation of the intensities in this experimental setting involves the
matrix elements of a rather complicated interaction Hamiltonian.\cite{MO69}
We settle for an estimate of the relative intensities of these transitions and consider
the simplest possible interaction operator (in fact, one of the terms appearing in the
full expression), which is proportional to $S_i^{(\pm)}$, where
$i$ denotes the central atom of the dynamic shell. The contribution of each
transition~\eqref{eq:trans} to the total intensity is proportional to  
\begin{equation}
    |\langle k, S^{(z)}=S_\text{N\'eel}\pm1
    |S_i^{(\pm)}|
    \text{G.S.},S^{(z)}=S_\text{N\'eel}\rangle|^2.
    \label{eq:intens}
\end{equation}
We have found that the main effect of the simplified transition matrix
element~\eqref{eq:intens} is to suppress the contributions of high-energy
transitions.  Note also that the combinatorial factor is already included
in our sampling of the simulation results. Finally, we add up the contributions
of all the possible transitions in our 1000 samples, calculate the convolution
of this result with the measured instrumental resolution function (a Gaussian with moderate
energy-dependent width), and normalize the result to match the maximum
count of the experimental curves (it would seem better to match the integrated
intensity in the experimental range, but as we will see, the resulting widths
are too narrow).

\begin{figure}
 \includegraphics[width=8cm]{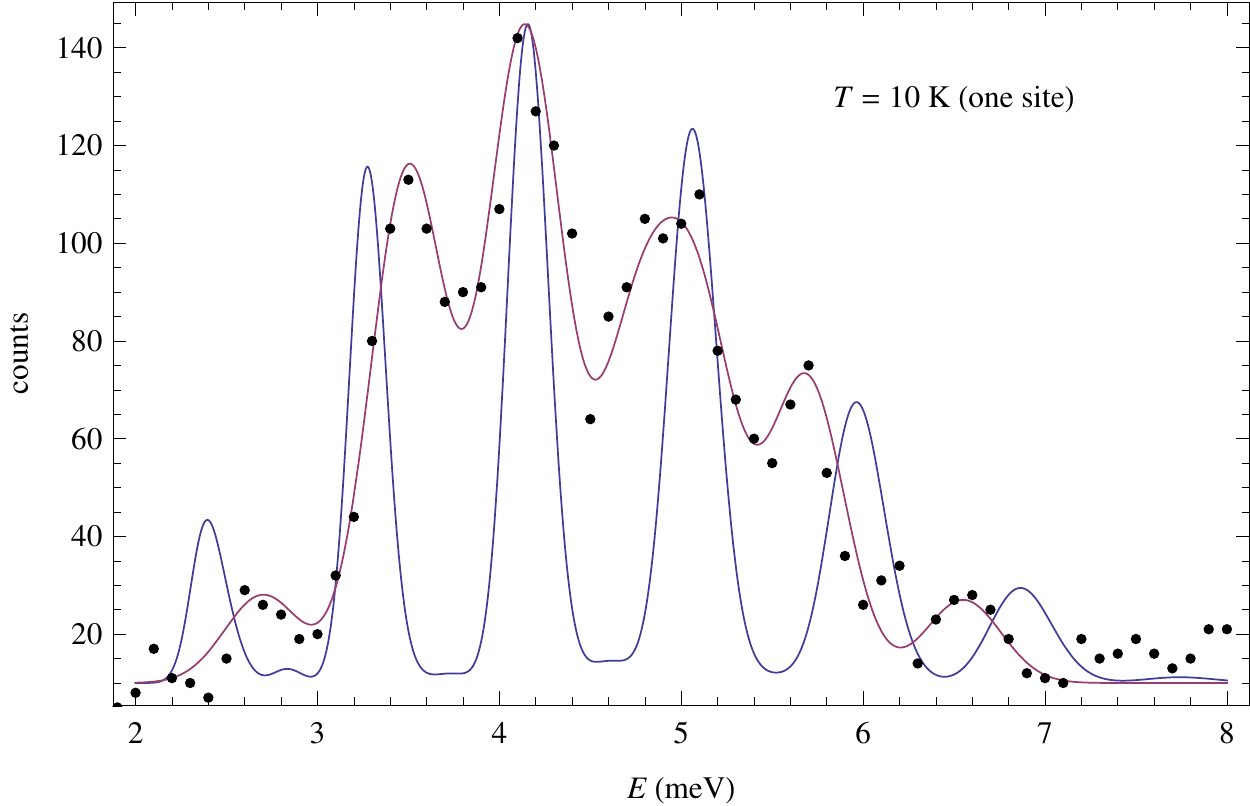}
 \includegraphics[width=8cm]{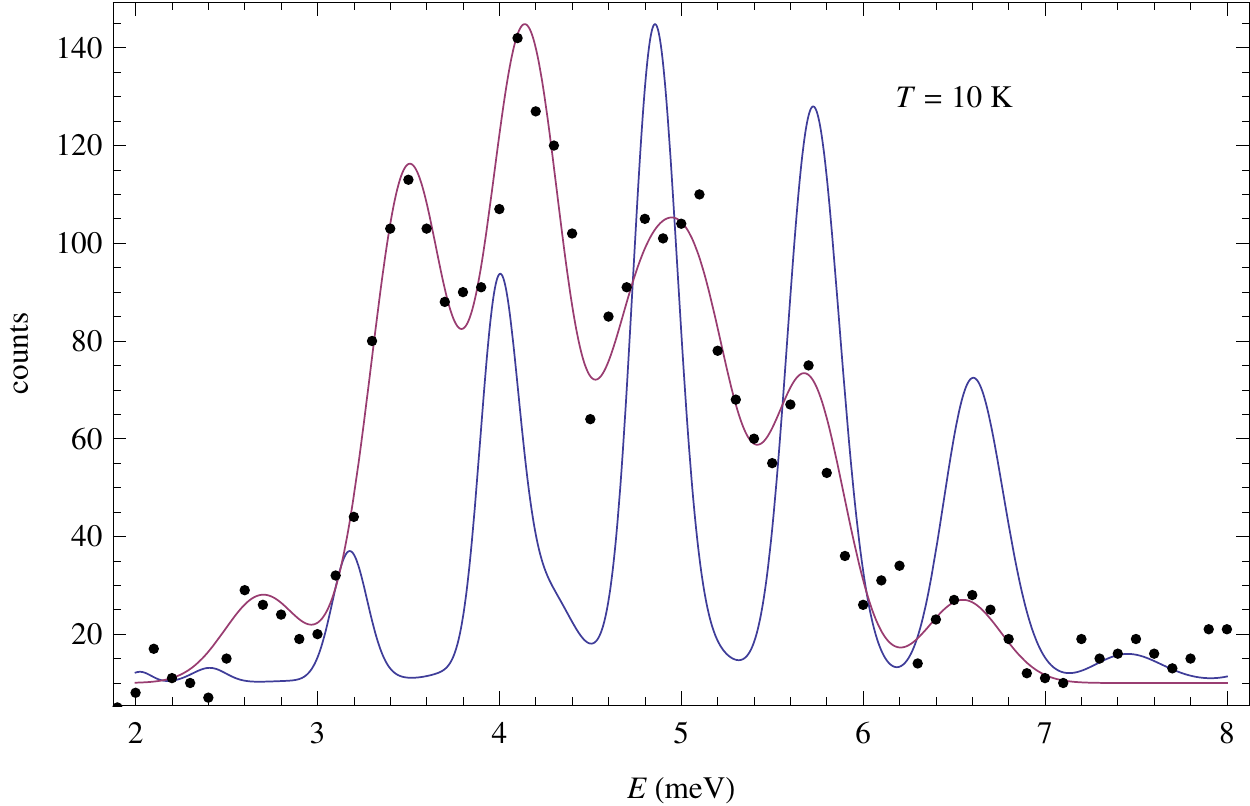}
 \caption{\label{stdparam} Experimental data for $q=0.5$,
numerical fitting and simulated spectrum
 at $T=10~\rm{K}$ with the parameters of the pure sample~\cite{HU70} for:
 (a) a one-site calculation corresponding to the dynamic shell of Fig.~\ref{unitcells}(a);
 (b) calculation corresponding to the dynamic shell of Fig.~\ref{unitcells}(b).}
\end{figure}
\begin{figure}
 \includegraphics[width=8cm]{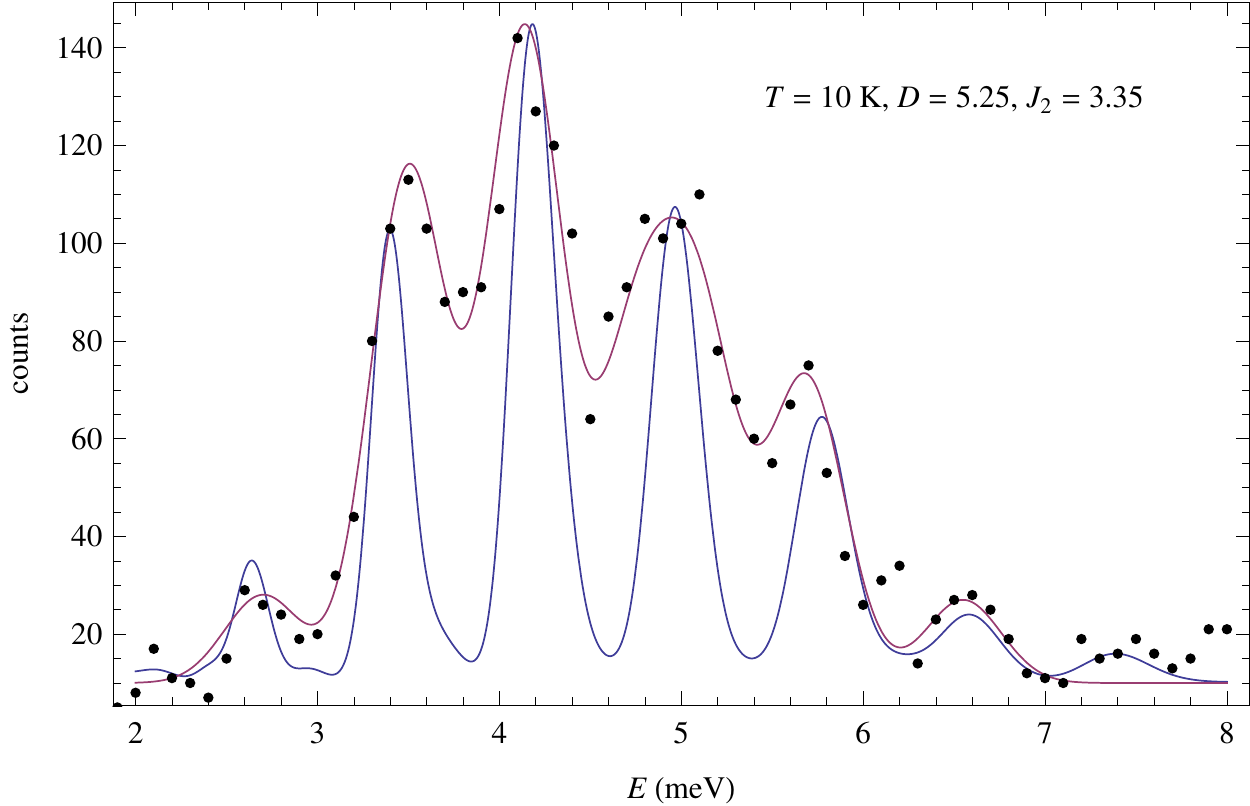}
 \includegraphics[width=8cm]{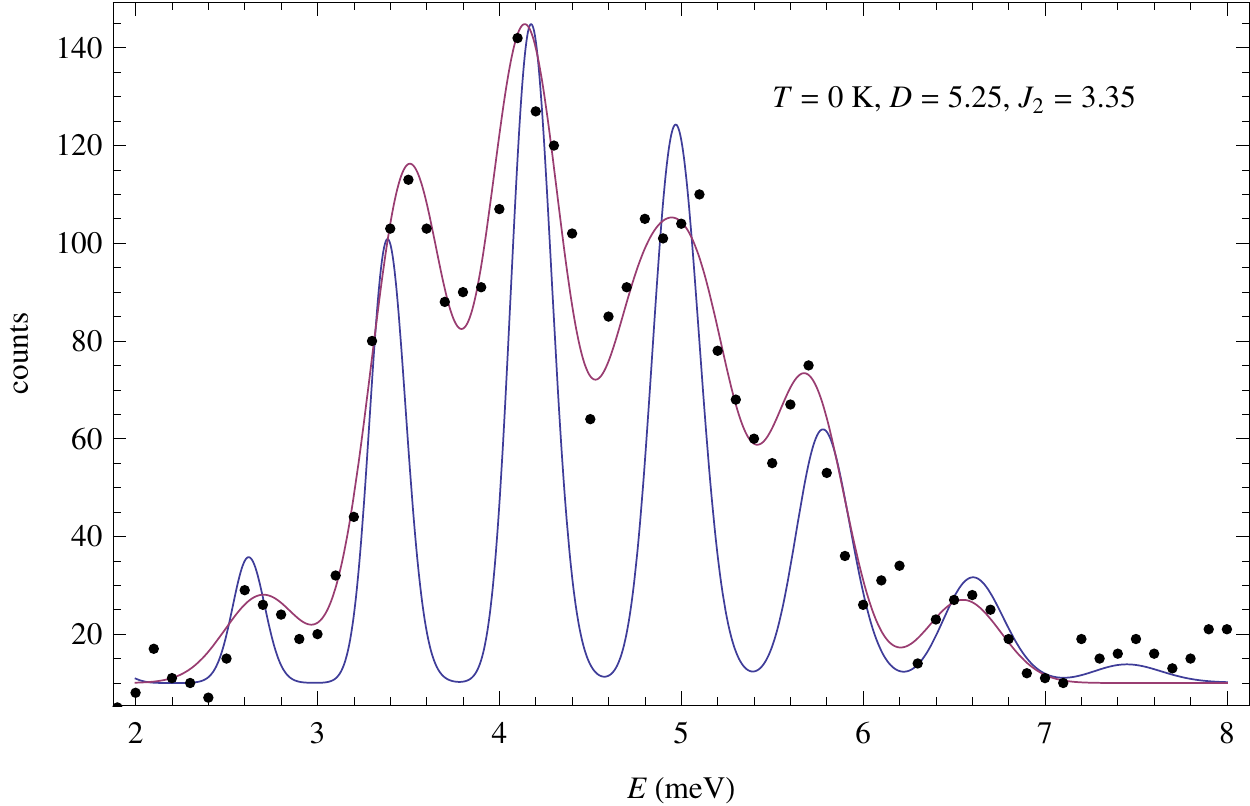}
 \caption{\label{bestparam}  Experimental points, numerical fitting and simulated spectrum
 with fitted values of $D$ and $J_2$ corresponding to the dynamic shell
 of Fig.~\ref{unitcells}(b) for: (a) $T=10~\rm{K}$; (b) $T=0~\rm{K}$ (i.e., purely antiferromagnetic
 environment).}
\end{figure}

Figure~\ref{stdparam} shows the results of these procedures for the two dynamic
shells illustrated in Fig.~\ref{unitcells} in the energy range between $2~\rm{meV}$ and
$8~\rm{meV}$. The figures show the experimental points, a numerical fitting to these
points, and our simulation results. The one-site calculation reproduces quite
well the main features of the experimental spectrum, including the average position
of the peaks (controlled by the anisotropy parameter $D$), although the separation
between peaks (controlled by the coupling constant $J_2$) is too large and, as we
anticipated, the widths are too narrow.

Although the simulation results corresponding to the dynamic shell of Fig.~\ref{unitcells}(b)
feature wider widths, the average position is clearly shifted to high energies, which suggests
that, not only the value of $J_2$, but also the value of the anisotropy parameter $D$ may be
too large in this context. A possible explanation might be related to the method by which
the pure sample parameters~\cite{HU70} are obtained, whereby a semiclassical approximation
is used to determine the spectrum parametrically as a function of $D$, $J_1$, $J_2$ and
$J_3$, and later these parameters are fitted to match the experimental results. In essence,
this procedure involves a calculation to first order in $1/S$, which should give better
results for $S\gg 1$ i.e., for the one-site approximation. This possibility has already been
noticed. For example, in Ref.~\onlinecite{PA94} certain empirical relations between the parameters
of the pure of the diluted sample are proposed.

Following these ideas, in Fig.~\ref{bestparam}(a) we show the result of an optimization
of the parameters $J_2$ and $D$ to match the experimental results, which yielded
$J_2=3.35~\text{cm}^{-1}$ and $D=5.25~\text{cm}^{-1}$. Unfortunately, the correlation
between these parameters and the uncertainties prevents a more accurate determination
of these values or the simultaneous optimization of the less significant $J_1$ and $J_3$
that we have kept fixed. Although the widths of the peaks are still too narrow, the intensities
are quite well accounted for, and even a last peak at $E\approx 7.4~\text{meV}$
seems to be reproduced.

Finally, as an estimate of the thermal effects in our simulations, in Fig.~\ref{bestparam}(b)
we show a similar calculation at $T=0~\rm{K}$, i.e., with a purely antiferromagnetic state
(no thermal disorder) of the environment. Note the distinctly narrower widths and
poorer intensity relations between the peaks, particularly at high energies.

There are a variety of possible reasons to explain the larger
experimental widths: our scattering operator is oversimplified, as it does not
take into account the relative orientation of the lattice and the wave vector
of the incoming neutron, which we also assume perfectly well defined (i.e., we neglect the
spread of the neutron beam); the dynamic shells and their environments have been obtained
from a classical (rather than quantum) Monte Carlo, which surely overestimates the spin
ordering at low temperatures; and the dynamic shells are limited to spins and
their immediate neighbors.

\section{Discussion}\label{sect:discussion}

In summary, we have presented high-resolution spectra from neutron scattering
experiments, conducted over $\rm{Fe}_x\rm{Zn}_{1-x}\rm{F}_2$ close to its
percolation threshold. We model these spectra in terms of a site diluted
Heisenberg model, containing both ferromagnetic and antiferromagnetic exchange
interactions. In spite of its simplicity and with only a moderate adjustment
of the parameters, the proposed model accounts quite well for the position and
intensity relations of the peaks in the spectra. This success is probably due
to the validity of our main hypothesis, namely the local nature of the spin
excitations in these systems which lie close to the percolation threshold for
the $\rm{Fe}$ lattice.

Whereas local spin excitations dominate
the energy spectrum for $x$ near
$x_{\mathrm p}$ in $\rm Fe_{1-x}Zn_xF_2$, we cannot rule out a
very small contribution from fracton excitations in a similar energy range.
Fracton excitations as well as local spin excitations
coexist in isotropic systems and both may exist in the
small anisotropy $\rm Mn_{1-x}Zn_xF_2$ system as $x$
approaches $x_{\mathrm p}$.  In that case, modeling the local spin
excitations could aid in separating the two types of
excitations, allowing the characterization of local spin
excitations as well as the persistence of fracton excitations
under conditions of weak anisotropy.


\subsection*{Acknowledgments}
We acknowledge partial financial support from MICINN, Spain, (Grant
Nos. FIS2009-12648-C03 and FIS2011-22566), and from UCM-Banco Santander
(GR32/10-A/910383, GR58/08-910556). V.M.-M. thanks the \emph{del Amo} foundation
and the hospitality of the Physics Department of U. California-Santa Cruz, where part of this work
was performed.  We thank the members of the Neutron Scattering
Laboratory, Institute for Solid State Physics, the University
of Tokyo for supporting our experiments.

\end{document}